\documentclass[aps,pra,amsfonts,nofootinbib,reprint,letterpaper,showpacs]{revtex4-1}

\usepackage{xspace}
\usepackage{graphicx}
\usepackage{dcolumn}
\usepackage{braket}
\usepackage{srcltx}
\usepackage{amsmath}

\newcommand{\figab}{Fig.~}
\newcommand{\tabab}{Table~}
\newcommand{\secab}{Sec.~}
\newcommand{\eqab}{Eq.~}
\newcommand{\citab}{Ref.~}

\newcommand{\abini}{\textit{ab~initio}\xspace}
\newcommand{\rmat}{\textsl{R}-matrix\xspace}
\newcommand{\otwo}{O$_2$\xspace}
\newcommand{\otwob}{\textbf{O}$_\mathbf{2}$\xspace}
\newcommand{\otwom}{O$_2^-$\xspace}
\newcommand{\ssigp}{$^1\Sigma_\mathrm{g}^+$\xspace}
\newcommand{\ssigum}{$^1\Sigma_\mathrm{u}^-$\xspace}

\newcommand{\sdelg}{$^1\Delta_\mathrm{g}$\xspace}
\newcommand{\dpig}{$^2\Pi_\mathrm{g}$\xspace}
\newcommand{\dpigb}{$\mathbf{^2\Pi_g}$\xspace}

\newcommand{\tpiu}{$^3\Pi_\mathrm{u}$\xspace}
\newcommand{\tsigm}{$^3\Sigma_\mathrm{g}^-$\xspace}
\newcommand{\tsigu}{$^3\Sigma_\mathrm{u}^+$\xspace}
\newcommand{\tsigum}{$^3\Sigma_\mathrm{u}^-$\xspace}
\newcommand{\odelg}{$^1\Delta_\mathrm{g}$\xspace}
\newcommand{\tdelu}{$^3\Delta_\mathrm{u}$\xspace}

\begin{document}
  \title{Low-energy Electron collisions with \otwob: Test of Molecular \rmat without Diagonalization}
  \author{\firstname{Michal} \surname{Tarana}}
  \email{michal.tarana@jila.colorado.edu}
  \author{\firstname{Chris} H. \surname{Greene}}
  \email{chgreene@purdue.edu}
  \affiliation{Department of Physics and JILA, University of Colorado,
  Boulder, Colorado 80309-0440, USA}
  \affiliation{Department of Physics, Purdue University, West Lafayette,
  Indiana 47907, USA}
  
  \begin{abstract}
    Electron collisions with \otwo at scattering energies below 1\thinspace eV are studied in the fixed-nuclei approximation for a range of internuclear separations using the \abini molecular \rmat method.
    The \dpig scattering eigenphases and quantum defects are calculated.
    The parameters of the resonance and the energy of the bound negative ion are then extracted.
    Different models of the target that employ molecular orbitals calculated for the neutral target are compared with models based on anionic orbitals.
    A model using a basis of anionic molecular orbitals yields physically correct results in good agreement with experiment.
    An alternative method of calculation of the \rmat is tested, where instead of performing a single complete diagonalization of the Hamiltonian matrix in the inner region, the system of linear equations is solved individually for every scattering energy.
    This approach is designed to handle problems where diagonalization of an extremely large Hamiltonian is numerically too demanding.
  \end{abstract}
  \maketitle
  \section{Introduction}
    After nitrogen, molecular oxygen is the most abundant molecule in the Earth's atmosphere.
    Therefore, its study is crucial to our understanding of planetary atmospheres, while also providing useful insight into the physics of gaseous discharges and laboratory and astrophysical plasmas.
    Theoretical and experimental research into electron interactions with \otwo has attracted significant scientific attention.
    Since a complete summary of the relevant references on this topic exceeds the scope of the present study (see the recent review by \citet{Itikawa2009} and references therein), we mention only those directly related to the problem addressed here.
    An experimental study of the resonant vibrational excitation of \otwo by the low-energy electron impact by \citet{Linder1971} shows oscillatory structures in the cross sections due to the \dpig resonance.
    \citet{Celotta1972} later measured the electron affinity of \otwo using molecular photodetachment spectroscopy.
    Those two complementary studies \cite{Linder1971,Celotta1972} provide a deeper understanding of the structure and dynamics of the \dpig state of \otwom \cite{Parlant1976}.
    
    Several previously published theoretical treatments based on the \rmat method deal with the \dpig resonance in the fixed-nuclei approximation \cite{Noble1986,Noble1992,Higgins1994,Higgins1995}.
    All those calculations yield the energy of the \dpig resonance at the equilibrium geometry of \otwo more than 0.6\thinspace eV above the value obtained from the experimental spectra \cite{Linder1971, Celotta1972,Teillet-Billy1987}.
    These studies are unable to reproduce the bound state of \otwom at larger internuclear separations without further adjustments of the \rmat poles \cite{Higgins1995,Noble1996}.
    \citet{Higgins1995} used the adjusted results of the \abini calculations to study resonant vibrational excitation by electron impact and compared the computed cross sections with the experimental work of \citet{Field1988}.
    The resonant vibrational excitation of \otwo at low energies has been revisited by \cite{Laporta2012}.
    Their theoretical treatment of the nuclear motion is based on the ``boomerang model'' and that utilized the same adjusted positions and widths of the resonances in the fixed-nuclei approximation calculated by \citet{Noble1996}.
    
    The energies of the \dpig resonance and the bound state of \otwom were later calculated more accurately by \citet{Ervin2003} using the methods of quantum chemistry designed originally for bound states.
    However, to our knowledge, no reliable \abini calculation exists that provides reliable scattering phase shifts at low energies ($<2$\thinspace eV) for different symmetries.
    
    Significant scientific attention has been paid to electron collisions with \otwo at energies above 2\thinspace eV as well.
    \citet{Teillet-Billy1987} found in their study based on an extended version of multichannel effective range theory that the \dpig resonance dominates the electronic excitation of the \sdelg and \ssigp states.
    This process was later studied experimentally \cite{Allan1995,Middleton1992,Middleton1994} and the results confirmed the existence of the resonances predicted by the previous calculations \cite{Noble1992,Noble1996,Higgins1994,Teillet-Billy1987}.
    
    Good correspondence between theoretical and experimental results for the excitation of the \sdelg and \ssigp electronic states by electron impact has been achieved (see paper \cite{Green2001} and references therein).
    However, the remaining discrepancies between the theoretical \cite{Teillet-Billy1989,Noble1992,Higgins1994} and experimental cross sections \cite{Teillet-Billy1989,Allan1995, Green2001} for excitation of the Herzberg pseudocontinuum (\ssigum, \tdelu, \tsigu) still constitute a challenge \cite{Green2001}.
    
    More recently, \citet{Tashiro2006a} presented another \rmat calculation of the elastic and electronically inelastic electron collisions with \otwo at energies above 5\thinspace eV that shows better agreement with the experimental results than previous \abini studies.
    
    The main goal of the research presented here is to develop an advanced \rmat calculation of electron collisions with \otwo in the fixed-nuclei approximation at scattering energies below 1\thinspace eV.
    In particular, improvements over previous calculations are sought to provide a more physical energy and width of the \dpig resonance and energy of the anionic bound state than was obtained in previous \rmat studies.
    The scattering eigenphases and the quantum defects discussed here can be employed in the investigation of the resonant vibrational excitation of \otwo based on the non-local resonance model \cite{Domcke1991} or energy-dependent vibrational frame transformation \cite{Gao1990}.
    The treatment of bound and continuum states of \otwom introduced here will be later adapted to study other symmetries of the anionic complex that are relevant to \otwom photodetachment and to the vibrational dynamics of that process.
    
    In comparison with previously published \abini studies, a more physical representation of the anionic complex electronic structure studies is achieved through the use of a more advanced treatment of electron correlation and polarization in the inner region and by employing a basis set of molecular orbitals optimized for the negative ion instead of the neutral target.  Note that in other contexts, this idea has also proven to be beneficial \cite{Schneider1979}.
    Current implementations of \abini electron-molecule scattering theory rely on expansions of the wave functions associated with both the neutral $N$-electron target and the $(N+1)$-electron system which use the same truncated basis of molecular orbitals.
    Although orbitals optimized for the neutral target are traditionally used in molecular \rmat calculations, it is far from obvious whether that is universally the most appropriate choice.
    The present study compares \rmat calculations performed using optimized neutral target molecular orbitals with those performed using optimized anionic orbitals, in order to ascertain which set is more appropriate.
    
    Another goal of this study is to test the feasibility of an alternative method of calculation of the \rmat that has the potential to treat larger basis sets and more extensive configuration interaction.
    The well-established approach preferred in the UK \rmat codes requires a single full diagonalization of the Hamiltonian matrix in the inner region \cite{tennyson-rev,Carr2012}.
    The \rmat is then easily calculated for any scattering energy, since its dependence on the energy is extremely simple in the eigenrepresentation of the Hamiltonian.
    The development of more efficient \rmat methods capable of treating molecules having a larger number of active electrons is one of the goals driving this research direction.
    The requirement of a full Hamiltonian diagonalization in the present implementation has made it quite challenging for the \rmat method to handle more complex polyatomic molecules and molecules for which electron correlation plays a key role.
    Evaluation of all the Hamiltonian matrix elements and their storing in the computer memory prior to the full diagonalization is necessary in the present implementation.
    That approach is computationally much more demanding than the more economical methods of the quantum chemistry of the bound states, where only a small fraction of the spectrum is calculated and the iterative methods of the diagonalization are routinely utilized. 
    Those require only evaluation the matrix elements that are used in the actual iteration.
    In addition, the complete set of the eigenvectors that have a dense structure is necessary in the present implementation of the \abini molecular \rmat method.
    That further raises issues with the memory limitations.

    The method presented in this study employs the solution of a system of linear equations having the same dimension as the Hamiltonian matrix, individually for every scattering energy.
    A number of other methods have previously been introduced to overcome this difficulty in solving problems requiring very large basis sets.
    The most promising among them is the partitioned \rmat \cite{Tennyson2004,Carr2012}.
    It consists of a single calculation of few lowest eigenvalues and eigenvectors of the Hamiltonian matrix and of a model-like approximation of the rest of the spectrum.
    While the partitioned \rmat method retains the advantage of a single diagonalization of a (reduced-dimensionality) Hamiltonian matrix, the alternative method presented here requires solution of the linear system of equations for every scattering energy.
    However, it is free of any model-like assumptions, and therefore, is a complementary approach to the partitioned \rmat method. It is hoped, however, that in combination with generalized quantum defect methods or multichannel effective range theory, the quantities to be calculated will be comparatively smooth functions of energy and can accordingly be calculated on a coarse energy mesh.
    Note that the method of calculation of the \rmat by solving the system of linear equations has been formulated previously by \citet{Collins1981}.
    Their linear-algebraic approach utilizes the static exchange approximation of the electron interactions in the inner region.

    The \abini theoretical description of electron collisions with \otwo at low scattering energies is challenging.
    Any successful treatment of the complicated polarization effects \cite{Hettema1994,Gonzalez-Luque1993} and electron correlation in the inner region requires a large basis set of configurations.
    Depending on the details of the model, the usual approach based on the full diagonalization of the inner-region Hamiltonian matrix \cite{tennyson-rev,Carr2012} can become intractable.
    That makes this system a suitable candidate to test the performance and limitations of the alternative method investigated here.
    As is discussed below, the linear solution method proves to be feasible and advantageous for calculations where the dimension of the Hamiltonian matrix in the inner region exceeds 40000.
    Several different models of the target are introduced below and, where possible, the performance of the traditional full diagonalization method is compared with that of the direct linear solution method.
    
    The rest of this paper is organized as follows: \secab\ref{sec:method} describes the method adopted to calculate the \rmat without any diagonalization, by direct solution of a linear inhomogeneous system of equations.
    Different models of the neutral target are discussed in \secab\ref{sec:targmodel} and the construction of the $(N+1)$-electron Hamiltonian matrix is described in \secab\ref{sec:scatcalc}.
    The results are presented in \secab\ref{sec:results} and the conclusions are summarized in \secab\ref{ref:conc}.
    
    Atomic units are used throughout, unless stated otherwise.
  \section{\rmat by linear equation solution (LES)}
    \label{sec:method}
    The idea of the \rmat method is the solution of the Schr\"odinger equation within a finite reaction volume $\Omega$ of the configuration space.
    The scattering properties of a many-particle system are known once the normal logarithmic derivative of the wave function and relevant surface amplitudes are specified on the surface $\Sigma$ enclosing the reaction volume.
    The goal of the theory is to determine this information in the form of an \rmat.
    The reaction volume in the molecular \abini \rmat method is specified by a sphere of radius $r_0$ chosen such that $r_i\leq r_0$, $r_i$ being the distance between the $i$th electron and the center of mass of the molecule.
    $r_0$ is large enough to contain all the complicated interactions of the electrons within the inner region, and sufficiently large that the target wave functions in the open and weakly-closed channels are negligible in the outer region.
    Interaction of the scattering electron with the target in the outer region is well approximated by the long-range multipole and dispersion potentials that couple different scattering channels \cite{tennyson-rev}.
    The total wave function can be written as \cite{Aymar1996}
    \begin{equation}
      \Psi_\beta(r,\omega)=\sum_{i=1}^{N_{ch}}\frac{1}{r}\Phi_i(\omega)F_{i\beta}(r),
      \label{eq:outexpand}
    \end{equation}
    where $r$ is the radial coordinate of the scattering electron and $\omega$ denotes all other spin-space coordinates of all the electrons (including the spin and angular coordinates of the scattering one).
    $\Phi_i(\omega)$ includes the electronic state of the target as well as the spherical harmonic of the scattering electron, $F_{i\beta}(r)$ is the scaled radial wave function of the scattering electron in the outer region ($r\geq r_0$) in the channel $i$. $N_{ch}$ denotes the total number of scattering channels retained (both open and closed), index $\beta=1,\dots,N_{ch}$ denotes different linearly independent solutions for the total energy $\epsilon$ of interest.
    If the radial derivative of $F_{i\beta}(r)$ is denoted as $F'_{i\beta}(r)$, then the \rmat on the surface $\Sigma$ is defined \cite{Aymar1996} as
    \begin{equation}
      \underline{R}= \left(\underline{F}\underline{F}'^{-1}\right)_{r_0}
      \label{eq:rmat}
    \end{equation}
    and is calculated by solving the Schr\"odinger equation in the inner region.
    Then it is used to match the solutions in the outer region and to calculate the $K$-matrix or other quantities that characterize the scattering process \cite{tennyson-rev, Aymar1996}.
    
    The approach to calculation of the \rmat tested in this work is based on the noniterative variational formulation of the \rmat method introduced by \citab\cite{Greene1983} that has been used extensively in a number of problems (see \citab\cite{Aymar1996} and references therein).
    Solutions of the time-independent Schr\"odinger equation at energy $\epsilon$ in the inner region obey
    \begin{equation}
      \hat{H}\Psi_\beta=\epsilon\Psi_\beta,
    \end{equation}
    where $\hat{H}$ is the electronic Hamiltonian of the $(N+1)$-electron system and $\epsilon$ is the total energy.
    $\Psi_\beta(r,\omega)$ can be expressed using a set of real basis functions $y_k(r,\omega)$ as $\Psi_\beta(r,\omega)=\sum_ky_k(r,\omega)C_{k\beta}$.
    Each of the basis functions $y_k$ can be expanded on $\Sigma$ as
    \begin{equation}
      y_k(r_0,\omega)=\sum_{i=1}^{N_{ch}}\frac{1}{r_0}\Phi_i(\omega)u_{ik}(r_0).
      \label{eq:bassurex}
    \end{equation}
    As is well known from \citet{Robicheaux1991} and \citet{Greene1988}, the \rmat can be expressed as
    \begin{equation}
      R_{ij}=\sum_{kk'}u_{ik}(\Gamma^{-1})_{kk'}u_{jk'},
      \label{eq:invrmat}
    \end{equation}
    where
    \begin{equation}
      \Gamma_{kk'}=2\int_{\Omega}y_k(\epsilon-\hat{H}-\hat{L})y_{k'}dV.
      \label{eq:gamdef}
    \end{equation}
    The integration is performed over the reaction volume $\Omega$.
    Notice that the matrix $\underline{\Gamma}$ is symmetric due to the presence of the Bloch operator $\hat{L}$ \cite{Bloch1957,Aymar1996,tennyson-rev}.
    The basis set used in the UK \rmat program suite allows for a close-coupling expansion of the total $(N+1)$-electron wave function in the inner region
    \begin{eqnarray}
      \Psi_\beta(r,\omega)=\hat{A}\sum_{i,k}C_{ik\beta}\frac{1}{r}\Phi_i(\omega)u_{ik}(r)\nonumber\\*
      +\sum_{p}D_{p\beta}\chi^{N+1}_p(r,\omega),
      \label{eq:molbasex}
    \end{eqnarray}
    where $u_{ik}(r)$ are the radial parts of the continuum orbital introduced to represent the scattering electron in the inner region and their values at $r_0$ are in general non-zero.
    The angular parts are included in $\Phi_i(\omega)$.
    The choice of the continuum orbitals depends on the symmetry of the target electronic states.
    These two are coupled to give the correct overall spin and spatial symmetry of $\Psi_\beta(r,\omega)$. Index $i$ denotes the scattering channel in the outer region and characterizes the electronic state of the target as well as the partial wave of the scattering electron.
    Therefore, each basis function that appears in the first sum in \eqab\eqref{eq:molbasex} has non-zero amplitude on the boundary $\Sigma$ and is associated with one scattering channel.
    Furthermore, the electrons must obey the Pauli principle and they are anti-symmetrized by operator $\hat{A}$. The second summation in \eqab\eqref{eq:molbasex} involves antisymmetric $(N+1)$-electron configurations $\chi_p^{N+1}$ that have zero amplitude on the boundary $\Sigma$ and where all the electrons occupy the orbitals associated with the target ($L^2$ configurations \cite{tennyson-rev}).
    Notice that in the terminology of the variational \rmat method \cite{Greene1983,Robicheaux1991,Aymar1996} the basis functions in the first sum in \eqab\eqref{eq:molbasex} correspond to the open part of the basis, while the second summation corresponds to the closed part.
    
    The eigenstates of the target and the $(N+1)$-electron basis functions are both expressed in terms of the complete active space configuration interaction (CAS CI) \cite{helgaker2000molecular}.
    It is worth mentioning at this point that the dimension of the ''closed`` part of the basis (second summation in \eqab\eqref{eq:molbasex}) is typically much larger than the dimension of the ''open`` part that has non-zero amplitudes on the boundary $\Sigma$.
    
    The modified Hamiltonian matrix for the inner region $\underline{H}_\Omega=\underline{H}+\underline{L}$ calculated using the basis expansion \eqref{eq:molbasex} is evaluated in the UK \rmat codes as well as the surface amplitudes $u_{ik}(r_0)$.
    The matrix $\underline{\Gamma}$ defined by \eqab\ref{eq:gamdef} can be easily calculated for each scattering energy $\epsilon$ of the interest and \eqab\eqref{eq:invrmat} can be used to calculate the \rmat.

    The product $\sum_{k'}\left\{\Gamma^{-1}\right\}_{kk'}u_{jk'}(r_0)$ is implemented as a solution of the linear system of equations, which is the most computationally demanding step of the calculation.
    
    A celebrated aspect of the approach to the calculation of the \rmat available in the UK \rmat suite nowadays is that the matrix $\underline{H}_\Omega$ is diagonalized only once and the energy dependence of the \rmat is calculated analytically \cite{tennyson-rev,Carr2012,Robicheaux1991}.
    However, the complete set of eigenvalues and eigenvectors in the given basis set is necessary for accurate evaluation of the \rmat using the expansion in the eigenstates.
    Beyond certain size of the basis set the matrix storage hits the memory limit or the time necessary to diagonalize the modified Hamiltonian becomes too long for practical calculations.

    The approach based on solution of the linear system for each individual energy of the interest becomes favorable in those cases, since both the time and memory required to solve one system of linear equations is significantly smaller than the time required to completely diagonalize a matrix of the same size.
    Furthermore, while existing computer routines for complete diagonalization are usually based on the full matrix storage, several modern computer implementations of state of the art linear solvers are based on more economical sparse storage schemes.
    Since $\underline{H}_\Omega$ is often a sparse matrix, the approach introduced above enables calculations that use larger CI models than the diagonalization-based method.
    A more efficient parallel implementation of the linear solvers than of the algorithms for the complete diagonalization makes the method presented above even more favorable for large-scale calculations performed on high-performance computer clusters.
    A calculation of the \rmat for a single value of the energy using \eqab\eqref{eq:invrmat} can be executed in the parallel mode in those environments and calculations for multiple energies can be trivially parallelized.
    
    Several other alternatives to the complete diagonalization of $\underline{H}_\Omega$ have been proposed and used \cite{Beyer2000,Tennyson2004,Carr2012,tennyson-rev,Tarana2007}.
    They are based on the accurate calculation of a modest number of the lowest eigenvalues and eigenvectors, while the rest of the spectrum is approximated using various models.
    These approximate approaches preserve the advantage of the single diagonalization and of the analytical energy-dependence of the \rmat.
    The most promising method among those is the partitioned \rmat \cite{Tennyson2004,Carr2012}.
    The method described above is not based on the diagonalization of $\underline{H}_\Omega$ and is free of any model-like assumptions.
  \section{Models of the Neutral Target}
    \label{sec:targmodel}
    All the \rmat calculations presented here were performed using the polyatomic UKRmol program suite \cite{tennyson-rev,Carr2012}, which
    uses a basis set of Gaussian-type orbitals (GTOs) in the inner region.
    The irreducible representations of the $D_{2h}$ point group are used, as this is the largest abelian subgroup of the true $D_{\infty h}$ symmetry point group of \otwo.
    The notation of the $D_{\infty h}$ point group is used throughout the rest of this paper, unless otherwise stated.
    
    The results are presented for a range of internuclear separations from $R_n=1.9$\thinspace a.u. to $R_n=3.5$\thinspace a.u., inside of which the potential energy curves for the ground electronic states of both \otwo (\tsigm) and \otwom (\dpig) reach their minima.
    The fixed-nuclei scattering phase shifts and smooth quantum defects obtained for that range are essential for future theoretical studies of resonant vibrational excitation by low-energy electrons and these entities also arise in the theoretical description of vibronic coupling in \otwom photodetachment.
    The molecular orbitals optimized using the state-averaged complete active space self-consistent field (SA-CASSCF) method implemented in the program package MOLPRO \cite{WK85,KW85} are employed in all the \rmat calculations presented here.
    
    The neutral target is represented by one of two different sets of the molecular orbitals.
    The first is calculated using an extensive GTO basis set of atomic natural orbitals (ANO) \cite{Widmark1990} and the SA-CASSCF molecular orbitals are optimized for the neutral molecule (three lowest electronic states are averaged with equal weights).
    This selection of the primary GTO basis set is motivated by Ref. \cite{Gonzalez-Luque1993}, where it was successfully used to calculate the electron affinity  of \otwo.
    Furthermore, \citet{Jones2010} found that this GTO basis is necessary to reproduce within CAS CI models the static dipole polarizabilities of various diatomic molecules containing oxygen, although the case of \otwo is not discussed in that study.
    Previously published \rmat studies \cite{Tashiro2006a,Higgins1994} suggest that polarization effects play an important role in the electron collisions with \otwo.
    The ANO basis set is expected to represent these effects more accurately than the more compact Gaussian basis sets employed by the previously published \rmat calculations.
    Two different models of neutral \otwo based on the ANO GTO basis and on the neutral molecular orbitals were tested.
    Two additional models of the target are introduced below, where Dunning's cc-pVTZ \cite{Dunning1989} GTO basis set is employed.
    The convergence of the scattering calculations with respect to the size of the CAS and the quality of representation of the correlation and polarization effects can be estimated from comparison of the scattering eigenphases and the energies of the stable negative ion calculated using these different models.
    
    The dominant valence electronic configuration of the ground state of \otwo is
    \begin{equation*}
      (2\sigma_g)^2(2\sigma_u)^2(3\sigma_g)^2(1\pi_u)^4(1\pi_g)^2.
    \end{equation*}
    It is natural to include the orbitals $3\sigma_u$, $2\pi_g$ and $2\pi_u$ in all the CAS models, since their occupation numbers in the ground state are larger than 0.01.
    All of the models presented here consist of 8 active valence electrons.
    The CAS models considered in the previously published \rmat studies \cite{Higgins1994, Noble1992, Tashiro2006} include all 12 valence electrons and a smaller set of the active orbitals than the CAS models introduced here.
    However, our preliminary tests suggest that the correlation effects due to the electrons $2\sigma_g$ and $2\sigma_u$ can be neglected for the range of the collision energies considered here.
    
    The first CAS model can be expressed as
    \begin{equation}
      (1\sigma_g2\sigma_g1\sigma_u2\sigma_u)^8(3\sigma_g3\sigma_u1\pi_u2\pi_u3\pi_u1\pi_g2\pi_g)^8.
      \label{eq:cas1}
    \end{equation}
    The \rmat calculations based on this CAS include the two energetically lowest target states from each irreducible representation that does not contribute to the static dipole polarizability of the ground state.
    \citet{Tashiro2006} suggested that the polarization effects were not sufficiently represented in previously published \rmat studies \cite{Noble1992,Higgins1994}.
    As a first step towards the improvement of this deficiency, the 30 lowest states from the irreducible representation $^3\Pi_u$ and the 42 lowest states from the irreducible representation $^3\Sigma_u^-$ (both have a non-zero dipole coupling with the ground state) are included in the expansion of the total wave function in the inner region.
    This set of target states and the CAS expressed by \eqab\eqref{eq:cas1} is denoted as Model 1 in the text below.
    
    In order to evaluate the convergence of the scattering calculations with respect to the number of the active molecular orbitals, another model of the target is introduced.
    The orbital $3\pi_u$ included in Model 1 is replaced by the orbital $4\sigma_g$.
    This CAS can be expressed as
    \begin{equation}
      (1\sigma_g2\sigma_g1\sigma_u2\sigma_u)^8(3\sigma_g4\sigma_g3\sigma_u1\pi_u2\pi_u1\pi_g2\pi_g)^8.
      \label{eq:cas24}
    \end{equation}
    The \rmat calculations based on this model include the 40 energetically lowest target states in every symmetry (both singlet and triplet spin states).
    This model is denoted as Model 2 in the text below.
    
    \begin{table}[htb]
      \caption{The number of molecular orbitals in every irreducible representation included in different CAS models of the target.
      The last line shows the number of continuum orbitals (COs) that is the same for all the \rmat calculations discussed here.
      The correspondence between the irreducible representations of the point groups $D_{2h}$ and $D_{\infty h}$ is shown as well.}
      \label{tab:orbitals}
      \begin{center}
        \begin{ruledtabular}
          \begin{tabular}{ccccccc}
            Symmetry ($D_{2h}$)&$a_g$&$b_{2u}$, $b_{3u}$&$b_{1g}$&$b_{1u}$&$b_{3g}$, $b_{2g}$&$a_u$\\
            Symmetry ($D_{\infty h}$)&$\sigma_g$, $\delta_g$&$\pi_u$&$\delta_g$&$\sigma_u$, $\delta_u$&$\pi_g$&$\delta_u$\\
            \hline
            Model 1&3&3&0&3&2&0\\
            Models 2,3&4&2&0&3&2&0\\
            Model 4&4&3&0&3&2&0\\
            COs&37&21&18&21&18&7
          \end{tabular}
        \end{ruledtabular}
      \end{center}
    \end{table}
    The number of molecular orbitals from every irreducible representation included in the treatment of the inner region for every CAS model introduced here is summarized in \tabab\ref{tab:orbitals}.
    
    The scattering eigenphases calculated using the target models introduced above provide insight into the role of the excited electronic states and higher molecular orbitals in electron collisions with \otwo.
    However, they do not yield physically correct energy of the \dpig resonance and fail to describe the bound state of \otwom.
    In order to solve this deficiency two additional target models (Models 3 and 4) based on different MOs are introduced.
    They employ the Dunning cc-pVTZ GTO basis \cite{Dunning1989} and the CASSCF molecular orbitals that are optimized for the \dpig ground electronic state of \otwom.
    The choice of the cc-pVTZ basis is motivated by the experience with the \rmat calculations based on Models 1 and 2 discussed below and by the intention to use the obtained results in the prospective calculations of the nuclear dynamics using the vibrational frame transformation method \cite{Gao1989,Gao1990}.
    The \rmat calculations based on the ANO GTO basis require a rather large \rmat sphere ($r_0=16$\thinspace a.u.), which causes numerical difficulties with the vibrational frame transformation method.
    Models 3 and 4 yield spatially more compact molecular orbitals that can be confined inside the sphere with the radius $r_0=10$\thinspace a.u.
    As is discussed below, the \rmat calculations based on Models 3 and 4 yield at the scattering energies below 1\thinspace eV more physical results than the calculations utilizing Models 1 and 2.
    It should be kept in mind that the molecular orbitals calculated for \otwom do not have any straightforward physical interpretation for the internuclear distances, where the anion is not bound.
    They are used as the basis functions in the CI expansion of the bound target states and in the expansion of the $(N+1)$-electron scattering wave function, where the correct boundary condition is guaranteed by the continuum orbitals and by the the Bloch operator defined on the boundary of the reaction volume.
    
    Model 3 consists of the same configurations as Model 2 (see \eqab\eqref{eq:cas24}) and 33 target states from every irreducible representation are used in the expansion of the wave function in the inner region.
    The most complex CAS model constructed in the present study (Model 4) allows the active electrons to occupy both orbitals $4\sigma_g$ and $3\pi_u$ along with $4\sigma_u$.
    It can be expressed as
    \begin{equation}
      (1\sigma_g2\sigma_g1\sigma_u2\sigma_u)^8(3\sigma_g4\sigma_g3\sigma_u4\sigma_u1\pi_u2\pi_u3\pi_u1\pi_g2\pi_g)^8
      \label{eq:cas3}
    \end{equation}
    and it is introduced to study the stability of the \rmat calculations based on the anionic molecular orbitals with respect to the extension of the CAS.
    The expansion of the wave function in the inner region includes the 33 lowest target states from every irreducible representation (singlet and triplet spin states) in the \rmat calculation using this model.
    The dimension of $\underline{H}_\Omega$ for this model is large and its full diagonalization, implemented as a standard method of calculation of the \rmat in the UK codes \cite{tennyson-rev, Carr2012} would be numerically very demanding.
    This suggests that Model 4 is a suitable candidate for demonstrating the alternative approach discussed in \secab\ref{sec:method} which employs no diagonalization.
    \begin{table}[htb]
      \caption{Energy of the \tsigm ground electronic state of \otwo (a.u.) and vertical excitation energies (eV) for the lowest few excited states.
      The excitation energies calculated using Models 1--4 are compared with the experimental values quoted in the reference \cite{Teillet-Billy1987}.
      All the values are for the internuclear separation $R_e(O_2)=2.3$\thinspace a.u.}
      \label{tab:vertens}
      \begin{center}
        \begin{ruledtabular}
          \begin{tabular}{@{}lddddd}
            &\multicolumn{2}{c}{ANO}&\multicolumn{2}{c}{cc-pVTZ}&\\
            \cline{2-3}\cline{4-5}
            \multicolumn{1}{l}{State}&\multicolumn{1}{c}{Model 1}&\multicolumn{1}{c}{Model 2}&\multicolumn{1}{c}{Model 3}&\multicolumn{1}{c}{Model 4}&\multicolumn{1}{c}{Ref. \cite{Teillet-Billy1987}}\\
            \hline
            $^3\Sigma_\mathrm{g}^-$&-149.8557&-149.8516&-149.8186&-149.8430&\\
            $^1\Delta_\mathrm{g}$&1.014&1.008&1.005&1.012&0.98\\
            $^1\Sigma_\mathrm{g}^+$&1.871&1.828&1.815&1.822&1.65\\
            $^1\Sigma_\mathrm{u}^-$&5.875&5.834&5.844&5.851&6.12\\
            $^3\Delta_\mathrm{u}$&6.141&6.094&6.099&6.124&6.27\\
            $^3\Sigma_\mathrm{u}^+$&6.303&6.254&6.252&6.282&6.47\\
            $^3\Sigma_\mathrm{u}^-$&9.458&9.503&9.477&9.322&9.25\\
            $^1\Delta_\mathrm{u}$&11.869&11.926&11.920&11.755&11.8\\
          \end{tabular}
        \end{ruledtabular}
      \end{center}
    \end{table}
    
    The energy of the \tsigm ground state of \otwo calculated for the equilibrium internuclear separation $R_e(\text{O}_2)=2.3$\thinspace a.u. is for Models 1--4 compared in \tabab\ref{tab:vertens}.
    Since in Models 1 and 2 the neutral target is represented using a larger ANO GTO basis and the molecular orbitals are optimized for the ground electronic state of the neutral molecule, it is not surprising that these models yield lower energy of the ground state than Models 3 and 4.
    On the other hand, the highest energy of the \otwo ground state obtained from Model 3 is a consequence of the smaller GTO basis set and of the fact that the wave function of the neutral target is expanded in the truncated basis set of the orbitals optimized for \otwom.
    
    The vertical excitation energies for the lowest eight electronic states calculated for Models 1--4 are also compared in \tabab\ref{tab:vertens}.
    In general, they are in good agreement with each other.
    Note that the excitation energy of the lowest excited state \odelg is for all the Models 1--4 close to the experimental value quoted in the reference \cite{Teillet-Billy1987}.
    The agreement with the experiment is less convincing for the higher excited states, although the correspondence between Models 1--4 is obvious.
    \tabab\ref{tab:vertens} as well as \tabab II in the reference \cite{Tashiro2006} suggests that the change of the primary GTO basis as well as the presence or absence of the orbitals $4\sigma_g$ and $3\pi_u$ in the CAS do not dramatically affect the vertical excitation energies.
    
    Since one of the goals of the present study is to provide the fixed-nuclei scattering eigenphases and energies of the anionic bound states for a future study of the vibrational dynamics, it is important to represent the target correctly also for different geometries than the equilibrium one.
    The vibrational energy $\omega_e(\text{O}_2)$ is a suitable quantity that suggests how well the different models introduced above represent the potential energy curve of the ground state near the equilibrium.
    In fact all the Models 1--4 yield values similar to each other and to the previously published experimental results \cite{Ervin2003}.
    Comparison of Model 1 and Model 3 with the experimental value is shown in \tabab\ref{tab:spectroscopic}.
    \begin{table*}[htb]
    \caption{The equilibrium internuclear distance $R_e$, vibrational frequency $\omega_e$ of the neutral target and of the \dpig anion and the electron affinity $E_A$. Calculations based on Model 1 and Model 3 are compared with the experimental results. The non-zero components of the tensor of the static dipole polarizability of the neutral target ($\alpha_{xx}$, $\alpha_{zz}$) at the equilibrium geometry calculated using the same models are compared with previously published theoretical values.}
      \label{tab:spectroscopic}
      \begin{center}
        \begin{ruledtabular}
          \begin{tabular}{@{}ldddddd}
            & \multicolumn{3}{c}{\otwo $(X^3\Sigma_\mathrm{g}^-)$} & \multicolumn{3}{c}{\otwom $(X^2\Pi_\mathrm{g})$}\\
            \cline{2-4}\cline{5-7}
            & \multicolumn{1}{c}{\text{Model 1}} & \multicolumn{1}{c}{\text{Model 3}} & \multicolumn{1}{c}{\text{Previously published}} & \multicolumn{1}{c}{\text{Model 1}} & \multicolumn{1}{c}{\text{Model 3}} & \multicolumn{1}{c}{\text{Experimental}}\\
            & & & \multicolumn{1}{c}{\text{values [Ref.]}} & & & \multicolumn{1}{c}{\text{values [Ref.]}}\\
            \hline
            $R_e$ (a.u.) & 2.30 & 2.32 & 2.28\text{ \cite{Ervin2003}} & 2.55 & 2.58 & 2.55\text{ \cite{Celotta1972}}\\
            $\omega_e$ (eV) & 0.192 & 0.185 & 0.196\text{ \cite{Ervin2003}} & 0.135 & 0.133 & 0.137\text{ \cite{Celotta1972}}\\
            $\alpha_{xx}$ (a.u.) & 0.08 & 0.22 & 7.12\text{ \cite{Hettema1994}}& & & \\
            $\alpha_{zz}$ (a.u.) & 9.31 & 9.2  & 13.51\text{ \cite{Hettema1994}}& & & \\
            $E_A$ (eV) & & & & -0.408 & 0.375 & 0.448\text{ \cite{Ervin2003}}
          \end{tabular}
        \end{ruledtabular}
      \end{center}
    \end{table*}
    
    As is discussed below, the polarization effects play important role in the scattering at low impact energies.
    It is their representation that requires rather large number of the target states included in the expansion of the scattering wave function in the inner region.
    Thus, it is interesting to calculate the static dipole polarizability of the ground state of the neutral molecule to estimate, how well different models introduced above represent these effects.
    The components of the tensor of the static dipole polarizability $\alpha$ of the ground state $\varphi_0$ can be calculated using the sum-over-states formula
    \begin{equation}
      \alpha_{rs}=\frac{1}{2}\sum_{k>0}\frac{\bra{\varphi_0}\hat{d}_r\ket{\varphi_k}\bra{\varphi_k}\hat{d}_s\ket{\varphi_0}}{\Delta E_k}\qquad r,s\in\{x,y,z\},
    \end{equation}
    where $\varphi_k$ are the excited states, $\hat{d}_{r,s}$ are the Cartesian components of the operators of the transition dipole moments and $\Delta E_k$ is the excitation energy from $\varphi_0$ to $\varphi_k$. The summation should be performed over the complete set of the eigenstates (only the states of the symmetry \tpiu or \tsigum have a non-zero contribution) including the continuum states.
    Here it is performed over all the target states included in the expansion of the scattering wave function in the inner region.
    The only non-vanishing components of $\alpha$ for the homonuclear diatomic molecule are the diagonal ones $\alpha_{zz}$ and $\alpha_{xx}=\alpha_{yy}$.
    One can see in \tabab\ref{tab:spectroscopic} that Models 1 and 3 yield a similar value of $\alpha_{zz}$ and it is more than 68\% of the value that was previously accurately calculated by \citet{Hettema1994}.
    On the other hand, both models yield significantly underestimated value of $\alpha_{xx}$ that is more than one order of magnitude lower than the accurate calculations of \citet{Hettema1994}.
    In fact, all the Models 1--4 introduced above yield very similar values of $\alpha$ and they show rather poor representation of the component that is perpendicular to the internuclear axis.
    It is worth mentioning that our test calculations (not published here) that utilize the pseudocontinuum orbitals in the CAS in addition to the molecular orbitals \cite{Gorfinkiel2005,tennyson-rev,tarana-tennyson,Jones2010} yield a higher value of $\alpha_{xx}$. However, the \rmat calculations with those huge basis sets are computationally too demanding to be practically performed (see also Appendix \ref{sec:aprmps}).
    \section{Scattering Calculations}
    \label{sec:scatcalc}
    The continuum basis for the inner region is constructed in the polyatomic UKRmol program using additional GTOs with the centers that coincide with the center of the \rmat sphere.
    A sufficient number of these GTOs is diffuse enough to have non-zero values on the boundary of the inner region.
    The exponents are optimized using the program GTOBAS \cite{Faure2002} and the resulting functions are orthogonalized on the set of the molecular orbitals.
    This procedure yields a set of continuum-type orbitals in the inner region.
    All the \rmat calculations presented here include the continuum orbitals with orbital angular momenta $l=0,1,2,3,4,5$.
    Their number in every irreducible representation is identical for all the Models 1--4 and is listed in \tabab\ref{tab:orbitals}.
    
    Models 1 and 2 are based on the ANO GTO basis that yields quite diffuse molecular orbitals.
    The corresponding \rmat calculations require quite a large sphere with radius $r_0=16$\thinspace a.u.
    A similarly large \rmat sphere was also necessary in the \rmat studies~\cite{tarana-tennyson,Tarana2009a} of electron collisions with other molecules having a sizable polarizability.
    The GTO basis set cc-pVTZ used in Models 3 and 4 yields target orbitals that are more compact and can be confined within the sphere of the radius $r_0=10$\thinspace a.u.
    
    The total number of target states (summed over all the irreducible representations and spin states) included in the expansion of the scattering wave function is listed in \tabab\ref{tab:laham} for Models 1--4 along with the dimensions of the corresponding Hamiltonian matrices $\underline{H}_\Omega$.
    \begin{table}[htb]
      \caption{The number of the target states $N_t$ included in the close-coupling expansion of the total wave function in the inner-region and the dimension of $\underline{H}_\Omega$ is compared for Models 1--4.
      $t_C$ is the CPU time in hours required for the calculation of all the matrix elements, $t_D$ is the CPU time in hours necessary to diagonalize $H_\Omega$ using the LAPACK subroutine DSYEVD and $t_L$ is the CPU time in hours necessary to calculate the \rmat for a single energy using the linear solver PARDISO.}
      \label{tab:laham}
      \begin{center}
        \begin{ruledtabular}
          \begin{tabular}{@{}cccccc}
            Model & $N_t$ & Dimension of $\underline{H}_\Omega$ & $t_C$ (h)& $t_D$ (h)& $t_L$ (h)\\
            \hline
            1 & 128 & 26168&17.7&6.4&0.6\\
            2 & 640 & 22456&3.25&3.9&0.92\\
            3 & 528 & 22512&4.5&4.17&1.2\\
            4 & 528 & 127012&97.7&&62.2
            \end{tabular}
        \end{ruledtabular}
      \end{center}
    \end{table}
    Quite a large number of target states are needed to achieve convergence of the scattering $K$-matrix and its eigenphases for all the models discussed here.
    That, in combination with the large number of the active electrons and orbitals, is the reason for the substantial amount of CPU time required for the evaluation of all the elements of $\underline{H}_\Omega$ (see the column $t_C$ in \tabab\ref{tab:laham}).
    Note that the CPU time required for the complete diagonalization of $\underline{H}_\Omega$ ($t_D$ in \tabab\ref{tab:laham}) is for Model 2 and 3 comparable with the CPU time necessary for the construction of $\underline{H}_\Omega$.
    The large size of the Hamiltonian matrix for Model 4 dramatically complicates the complete diagonalization of the Hamiltonian matrix and for this case only the method based on solving the linear system of equations was applied.
    The CPU time required for the solution of the linear system of equations is smaller than the time necessary for the complete diagonalization (see the column $t_L$ in \tabab\ref{tab:laham}) for all models where both methods can be compared.
    However, it must be kept in mind that the linear system has to be solved for every scattering energy of interest.
    This makes the formulation of the outer-region problem in terms of the analytical quantum defects more favorable, since a smooth dependence on the energy permits the \rmat calculations to be performed for a smaller number of energy grid points followed by interpolation when possible.  Also, it should be kept in mind that the repetitive linear equation solution required at different energies is trivially parallelizable.
    Although the CPU time necessary to calculate the \rmat for Model 4 using the linear solver is rather high, the efficient parallelization allows for calculating the \rmat for a single value of the scattering energy on a computer with 8 CPUs in less than 12 hours.
    The large CPU time required for the construction of $\underline{H}_\Omega$ in Model 4 suggests that this is the most time-consuming step of the \rmat calculation independently of the complete diagonalization. 
    
    The large number of $(N+1)$-electron wave functions needed to achieve convergence for all of the Models 1--4 considered here is in fact the usual complication of the \abini calculations of the electron collisions with molecules with large polarizability \cite{Gil1994}.
    A useful computational method developed to treat this situation efficiently is the \rmat with pseudostates (RMPS) \cite{Gorfinkiel2005,tennyson-rev}, where the large number of true electronic states of the target is replaced by a smaller set of pseudostates \cite{tarana-tennyson}.
    We comment, however, that the straightforward application of this approach to the present problem has  not simplified the calculations and it does not improve the results.
    Further details are discussed in Appendix \ref{sec:aprmps}.
    
    The extensive number of the target states considered in the inner region problem can lead to very time consuming \rmat propagation in the outer region, if all the scattering channels are included in the outer-region problem (\eqab\eqref{eq:outexpand}) as well.
    This situation is similar to the RMPS calculations, where the \rmat propagation in the outer region usually requires more CPU time than the complete diagonalization of $\underline{H}_\Omega$ \cite{Gorfinkiel2005,tarana-tennyson}.
    Fortunately, the treatment of the wave function in the outer region can be simplified.
    As one can see in \tabab\ref{tab:vertens}, the threshold energy of the lowest electronically excited channel (\sdelg) is at 1\thinspace eV above the ground state of \otwo.
    That is the upper limit of the energy range considered in this study.
    All the scattering channels associated with the excited target states are strongly closed and although they play an important role in the inner region, they can be safely neglected in the outer region.

    Since previous \rmat studies \cite{tarana-tennyson,Tarana2009a} of the polarization effects suggest that their representation in the inner region plays a more important role than the polarization potential in the outer region, the \rmat propagation in the outer region is skipped in all the calculations presented here and the \rmat calculated at $r_0$ is used to match directly to the regular and irregular radial wave functions of the free particle in the outer region.
    Their linear combination determines the $K$-matrix and the scattering eigenphases.
    The calculations including different numbers of partial waves show that the \dpig electronic state of \otwom for the incident electron energies below 1\thinspace eV can be sufficiently well-represented in the outer region by the single partial wave $d$, i.e. $l=2$.
    Therefore, the problem in the outer region problem can be reduced to a single scattering channel.
    It is worth mentioning at this point that neglecting the long-range interaction of the target with the incident electron in the outer region that is predominated by the polarization potential, leads to a modification of the threshold behavior of the phase shift.
    While the dependence of the phase shift on the momentum of the incident electron $k$ is is in the present results $\eta(k)\propto k^5$ for $k\to 0$, if the potential $\propto r^{-4}$ was taken into account in the outer region, this would not be the leading term in the effective range expansion \cite{OMalley1961}.
    
    The corresponding phase shift $\eta(E)$ for the internuclear separations $R_n$, where the electronic state \otwom (\dpig) is not bound, can be parametrized by the Breit-Wigner formula
    \begin{equation}
      \eta(E)=\tan^{-1}\left(\frac{\Gamma/2}{E_r-E}\right)+\eta_0(E),
      \label{eq:bwform}
    \end{equation}
    where $\Gamma$ and $E_r$ are the width and position of the resonance, respectively and $E=k^2/2$ is the kinetic energy of the incident electron.
    The background phase shift $\eta_0(E)$ is a smooth function of the energy and it can be parametrized by a low-order polynomial.  This Breit-Wigner fitting formula is only valid for resonance energies well-separated from threshold, i.e. by many widths $\Gamma$.
    
    One of the goals of the present study is to provide the fixed-nuclei data required for the future calculations of the resonant vibrational excitation of \otwo and photodetachment of \otwom.
    To this end the analytical quantum defects $\mu_0(E)$ are calculated as a function of the scattering energy for a set of the internuclear separations.
    The analytical quantum defect $\mu_0(E)$ is a scattering quantity similar to the scattering eigenphase $\eta(E)$ in the sense that it specifies the linear combination of the general solutions of the Schr\"odinger equation in the outer region matching the boundary condition \eqref{eq:rmat}.
    These solutions are rescaled to remove the Wigner threshold factors and the analytical quantum defect is a function of the scattering energy that is smooth across the thresholds and it is well defined for both open and closed channels \cite{Greene1979,Greene1979a}.
    The smooth character of the analytical quantum defects even in the vicinity of the resonance makes this parametrization of the scattering wave functions particularly favorable in the context of the energy-dependent vibrational frame transformation \cite{Gao1989,Gao1990}.
    
    The reduction of the problem in the outer region to a single scattering channel and to a single partial wave $l=2$ (for incident electron energies below 1\thinspace eV) implies that the expansion \eqref{eq:outexpand} reduces to a single term.
    The scaled radial wave function of the scattering electron can be for $r\geq r_0$ parametrized as
    \begin{eqnarray}
      F_{11}(r)=\mathcal{N}\left[f^0_l(k,r)\cos(\pi\mu^0)\right.\nonumber\\*
      \left.-g^0_l(k,r)\sin(\pi\mu^0)\right]_{r\geq r_0}.
    \end{eqnarray}
    $f^0_l(k,r)=\sqrt{2/\pi}k^{-l}rj_l(kr)$ is the regular scaled radial wave function of the free particle, $g^0_l(r)=\sqrt{2/\pi}k^{l+1}rn_l(kr)$ is the irregular scaled radial wave function, $k$ is the momentum of the incident electron, $j_l$ and $n_l$ are the regular and irregular spherical Bessel functions, respectively and $\mathcal{N}$ is the normalization factor.

    Formulation of the problem in the outer region in terms of the smooth quantum defects $\mu^0$ is also favorable for the calculations of the anionic bound states in the fixed-nuclei approximation.
    For the range of nuclear geometries, where the anionic state is bound, its energy $E_b$ can be calculated by solving the equation \cite{Greene1979}
    \begin{equation}
      \pi\mu^0(E_b)=\pi-\arctan\left[(-2E_b)^{l+\frac{1}{2}}\right]\qquad E_b<0,
      \label{eq:boundener}
    \end{equation}
    where the single partial wave and single target electronic state is assumed.
    The advantage of this method compared to the widely used matching of the \rmat to the spherical Hankel functions (see the review \cite{tennyson-rev} and references therein) is that both sides of \eqab\eqref{eq:boundener} are usually smooth functions of energy and the complications due to the poles of the \rmat can be avoided.
  \section{Results}
    \label{sec:results}
    \subsection{The \dpigb Resonance at Equilibrium Geometry of the Neutral Target}
\begin{figure}
      \includegraphics[scale=0.68]{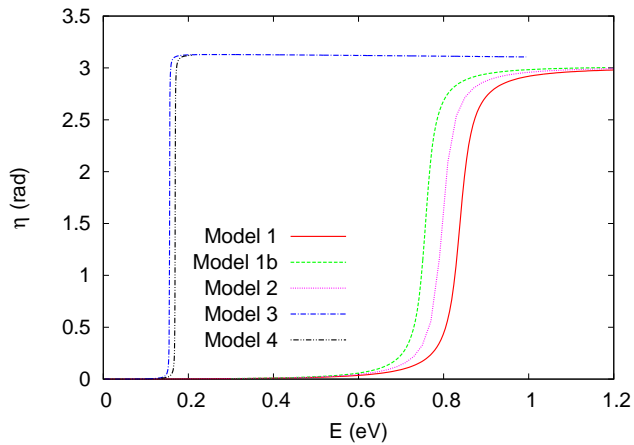}
      \caption{(Color online) The \dpig phase shift as a function of the incident electron energy calculated for the internuclear separation $R_e(O_2)=2.3$\thinspace a.u.
      Results for the Models 1--4 are compared.}
      \label{fig:eqphsum}
    \end{figure}

    \figab\ref{fig:eqphsum} shows the \dpig phase shift for the equilibrium internuclear distance of the neutral target $R_e(\text{O}_2)=2.3$\thinspace a.u. calculated using the
    models introduced in \secab\ref{sec:targmodel}.
    All the curves clearly show a narrow resonance with relatively small background.
    Model 1 yields the highest resonance position among all ($E_r=0.838$\thinspace eV).
    The expansion of the wave function in the inner region includes the 30 lowest target states from the irreducible representation \tpiu and the 42 lowest target state from the irreducible representation \tsigum, since those contribute to the polarizability of the ground state of the target.
    Only the two lowest target states are included from each of the other irreducible representations that have zero dipole coupling with the ground state of the target.
    
    Further tests showed that the higher target states from these irreducible representations are also important for achieving convergence of the phase shift within the space of configurations \eqref{eq:cas1}.
    The phase shift calculated including the 33 energetically lowest target states from every symmetry in the inner region is plotted in \figab\ref{fig:eqphsum} and denoted as Model 1b. 
    Inclusion of these additional states decreases the resonance position by $0.081$\thinspace eV to $E_r=0.757$\thinspace eV and adding even more excited states does not considerably change this value.
    
    The representation of the polarization effects by the CAS CI model, and the question of how well it is characterized by the static dipole polarizability of the target, has been the subject of several studies \cite{Gorfinkiel2005,tarana-tennyson,tennyson-rev}.
    The difference between the phase shifts denoted as Model 1 and Model 1b in \figab\ref{fig:eqphsum} demonstrates that the static dipole polarizability is not a sufficient measure of the polarization effects in the electron collisions with \otwo at low energies. 
    The decrease of $E_r$ with an increasing number of excited target states was also reported by \citet{Tashiro2006}, although that model takes into account only a significantly smaller set of target states; achieving convergence at low energies was not the goal of that study, in any case.
    
    In Model 2, the electrons can occupy the orbital 4$\sigma_g$ instead of the orbital 3$\pi_u$ included in Model 1.
    Although the expansion of the wave function in the inner region includes 40 excited states from every irreducible representation, the phase shift shows the resonance at slightly higher collision energy ($E_r=0.797$\thinspace eV) than Model 1b, as one can see in \figab\ref{fig:eqphsum}.
%
%
    
    In general, the comparison of the phase shifts calculated using Models 1 and 2 suggests that presence or absence of the molecular orbitals $4\sigma_g$ and $3\pi_u$ in the CAS does not dramatically affect the parameters of the resonance.
    The energy of the \dpig resonance calculated for $R_e(\text{O}_2)$ using Models 1 and 2 agrees well with the previously published \rmat calculations by \citet{Higgins1994} ($E_r=0.754$\thinspace eV).
    The CAS used in that study is smaller than in Models 1 and 3 and only the eight energetically lowest target states are included in the expansion of the wave function in the inner region.
    Therefore, none of the target states with $\Pi_g$ or $\Pi_u$ symmetry are included in that study (see \tabab\ref{tab:vertens}), although these states have significant contribution to the polarization effects. The absence of the higher molecular orbitals and excited target states is compensated by the virtual orbital that can be singly occupied only in the $(N+1)$-electron wave function \cite{Higgins1994} (the $L^2$ configurations in the close-coupling expansion \cite{tennyson-rev} or in \eqab\eqref{eq:molbasex}).
    The use of virtual orbitals raises the issues with the unbalanced treatment of the correlation in the target and in the scattering complex. That can lead to an ambiguity in the energy of the resonance.
    The models introduced in \secab\ref{sec:targmodel} are free of this complication.
    
    The resonance energy $E_r$ and width $\Gamma$ calculated for the internuclear separation $R_e(\text{O}_2)$ are summarized in \tabab\ref{tab:compares}.
    \begin{table}[htb]
      \caption{Energy $E_r$ and width $\Gamma$ of the \dpig resonance in \otwom at the equilibrium internuclear distance of the neutral target $R_e(\text{O}_2)$.
      Results of the molecular \rmat calculations using Models 1--4 are compared with the results in the previously published references.}
      \label{tab:compares}
      \begin{center}
        \begin{ruledtabular}
          \begin{tabular}{@{}ldd}
          &\multicolumn{1}{c}{$E_r$ (eV)}&\multicolumn{1}{c}{$\Gamma$  (eV)}\\
            \hline
            Model 1 &0.838&0.043\\
            Model 1b &0.757&0.034\\
            Model 2 &0.797&0.038\\
            Model 3 &0.154&3.6\times10^{-5}\\
            Model 4 &0.169&1.1\times10^{-3}\\
            \citet{Higgins1994}&0.754&0.031\\
            \citet{Noble1992}&0.700&0.026\\
            Derived from experiment \cite{Teillet-Billy1987}&0.090&8.5\times 10^{-5}
          \end{tabular}
        \end{ruledtabular}
      \end{center}
    \end{table}
    It shows that Models 1 and 2 yield a resonance width $\Gamma$ having the same order of magnitude.
    These values agree well with the results previously published by \citet{Higgins1994} and by \citet{Noble1992}.
    Our testing \rmat calculations performed with even larger CAS than those of Models 1 and 2 (not published here) show that the position and width of the \dpig resonance does not considerably change, when both orbitals $3\pi_u$ and $4\sigma_g$ are included in the CAS.
    
    Although the results obtained using Models 1 and 2 presented in \figab\ref{fig:eqphsum} and \tabab\ref{tab:compares} exhibit encouraging agreement with the previously published \rmat calculations~\cite{Higgins1994,Noble1992}, they do not agree well with the experimental study by~\citet{Linder1971} that found the energy of the \dpig resonance to be $E_r\approx0.1$\thinspace eV.
    A similar value was calculated by \citet{Ervin2003} using the stabilization method.
    This suggests that the relatively good agreement of the phase shifts calculated using Models 1 and 2 points towards the very slow convergence with respect to the number of the molecular orbitals included in the CAS.
    The authors of the previously published \abini studies \cite{Noble1992,Higgins1994,Tashiro2006} attribute the discrepancy between the theoretical and experimental results at energies below 1\thinspace eV to the insufficient treatment of the polarization effects, particularly in the outer region.
    Models 1 and 2 discussed above employ a quite extensive GTO basis set (ANO) that includes a subset of diffuse functions designed to represent the polarization effects.
    It is reasonable to expect that Models 1 and 2 account for these effects more than in any previously published \rmat study and the rather large \rmat sphere ($r_0=16\thinspace$ a.u.) should hold most of these effects in the inner region.
    However, even this improved treatment is not sufficient to provide more physical parameters of the \dpig resonance.
    
    The results suggest that the CAS CI representation employing the SA-CASSCF orbitals optimized for the neutral \otwo is not sufficient for the reliable scattering calculations in the energy range considered here.
    This conclusion is understandable in view of the previously published \abini studies of the adiabatic electron affinity $E_A$ of \otwo.
    It is another essential quantity that characterizes the \dpig state of \otwom at the equilibrium internuclear distance of bound \otwom ($R_e(\mathrm{O}_2^-)\approx2.6$\thinspace a.u. \cite{Ervin2003}).
    \citet{Gonzalez-Luque1993} compared the value of $E_A$ calculated using different CI models with the experimental value and found that the methods based on the CAS approach do not yield even a correct sign of $E_A$.
    According to that study, really extensive multi-reference configuration interaction (MRCI) calculations are required to obtain a value of $E_A$ comparable with the experimental results.
    Furthermore, \citet{Stampfuss2003} studied the contributions of the single and double excitations of the reference Hartree-Fock determinants to the binding energy of \otwom (\dpig) and compared it with the contribution of the triple and quadruple excitations.
    Their results show that both contributions are comparable.
    Therefore, the set of configurations, where several electrons are excited into the lowest few molecular orbitals (included in the CAS models introduced here), is not sufficient to yield the correct value of $E_A$.
    That requires inclusion of configurations where at least one electron occupies one of the higher molecular orbitals with an orbital number up to 12 in every irreducible representation.
    These orbitals are not included in the CAS models employed in this study and their further extension would lead to an extremely demanding construction of the Hamiltonian matrix $\underline{H}_\Omega$.
    
    Since the equilibrium internuclear separation of \otwo is only 0.3\thinspace a.u. smaller than the equilibrium geometry of \otwom, the same mechanisms are responsible for the slow convergence of the resonance energy $E_r$ at the equilibrium geometry of \otwo.
    
    In order to improve the results provided by Models 1 and 2, the CAS Model 3 has been introduced.
    It employs the SA-CAS MCSCF molecular orbitals of \otwom (\dpig), as is described above.
    The corresponding phase shift calculated for the internuclear separation $R_e(\text{O}_2)=2.3$\thinspace a.u. is also plotted in \figab\ref{fig:eqphsum}.
    It clearly shows a sharp resonance with a relatively smooth background.
    Fitting to the Breit-Wigner formula \eqref{eq:bwform} yields $E_r=0.154$\thinspace eV.
    This value agrees with the experimental results~\cite{Linder1971} much better than Models 1 and 2 (see \tabab\ref{tab:compares}).
    The calculation in the inner region includes 33 lowest target states from every irreducible representation and further increase of that number does not considerably change the phase shift.
    Interpretation of the improvement due to replacement of the orbitals optimized for the neutral target by the anionic molecular orbitals is not straightforward.
    In general, the orbitals optimized for the anion have more diffuse character than those calculated for the neutral target.
    Therefore, it is reasonable to expect that they are more suitable to represent the polarization effects than the molecular orbitals of the neutral target.
    This can partially compensate the absence of the higher orbitals in the CAS models mentioned above.
    On the other hand, the orbitals calculated for the neutral target are more suitable to represent the ground and excited states of the target than the anionic orbitals used in Model 3.
    It is possible that the lower energy of the resonance calculated using Model 3 is not only a consequence of the improved treatment of the interaction between the target and the scattering electron, but to some extent also an artifact of less accurate model of the target.
    In other words, part of the reason of the lower resonance energy in Model 3 is the increase of the target ground state energy with respect to Models 1 and 2 (see \tabab\ref{tab:vertens}).
    The CAS CI expansions of the target states and of the $(N+1)$-electron wave function in the same truncated set of the molecular orbitals show different convergence with respect to the number of orbitals and this convergence depends on their character.
    This general complication of the \abini \rmat calculations does not have a universal solution and the particular choice of the molecular orbitals apparently cannot yet be automated, but rather needs to be physically motivated. 
    
    The CAS Model 3 yields the vertical excitation energies of the target and the energy of the resonance that are in reasonable agreement with the experimental values (see \tabab\ref{tab:vertens} and \tabab\ref{tab:compares}).
    It should be emphasized that these results are not adjusted by an artificial overcorrelation of the $(N+1)$-electron system by introducing virtual orbitals~\cite{tennyson-rev, Noble1992, Higgins1994, Tashiro2006}.
    
    The CAS Model 4 allows the electrons to occupy both $4\sigma_g$ and $3\pi_u$ molecular orbitals that are included separately in different CAS models introduced previously.
    In addition, it includes the $4\sigma_u$ molecular orbital.
    This model yields the largest Hamiltonian matrix $\underline{H}_\Omega$ among all the Models 1--4 (see \tabab\ref{tab:laham}).
    It is introduced to study the stability of the \rmat calculations with respect to extending the CAS.
    The scattering phase shift (plotted in \figab\ref{fig:eqphsum}) shows encouraging agreement of the resonance position and width with Model 3, although the energy of the resonance $E_r=0.169$\thinspace eV is 15\thinspace meV higher than the value obtained from Model 3.
    This slight shift towards higher energies suggests that the molecular orbitals $3\pi_u$ and $4\sigma_u$ contribute more to the correlation of the target ground state than to the correlation of the $(N+1)$-electron system.
    The energy of the ground state of the target calculated using Model 4 is 0.66\thinspace eV lower than the energy calculated using Model 3 (see \tabab\ref{tab:vertens}).
    The good agreement between the $E_r$ calculated using Models 3 and 4 suggests that the improvement of the scattering results by employing the anionic molecular orbitals instead of those optimized for the neutral is not purely an artifact due to making the representation of the target worse.
    The phase shift is rather stable with respect to changes of the CAS that yields different energies of the target states.
    
    Since the construction of the Hamiltonian matrix $\underline{H}_\Omega$ in the \rmat calculations based on the most complex CAS Model 4 is computationally quite demanding (see \tabab\ref{tab:laham}), this was performed only for the equilibrium internuclear distance of the neutral target.
    Since those results show good agreement with the smaller Model 3, that more economical model was used to study the dependence of the \dpig bound and continuum states of \otwom on the internuclear distance discussed below.
    
    The effects of the vibrational nuclear dynamics prevent us from directly comparing the cross sections calculated in the fixed-nuclei approximation with experimental results for the energies of the scattering electron below 1\thinspace eV.
    Existence of the bound anionic state (discussed below) leads to well pronounced boomerang oscillations in the elastic scattering cross sections \cite{Linder1971,Higgins1995,Laporta2012} that do not appear in the fixed-nuclei calculations.
    Further theoretical study of these effects using the energy-dependent vibrational frame transformation \cite{Gao1989,Gao1990} based on the results presented here will be a subject of the forthcoming research.

    \subsection{The bound and continuum state of \otwom (\dpigb)}
\begin{figure}[htb]
      \includegraphics[scale=0.62]{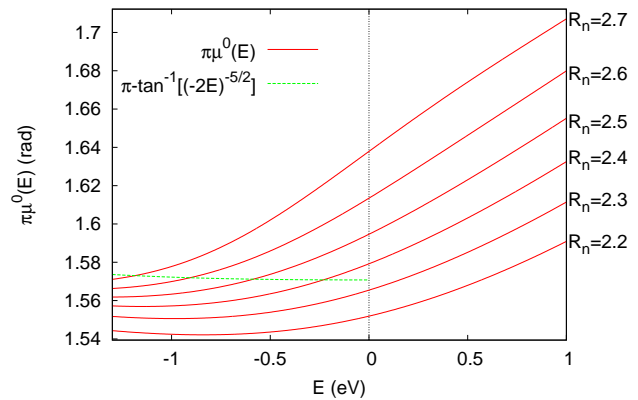}
      \caption{(Color online) Analytical quantum defects $\mu^0$ (solid lines) calculated using Model 3 as a function of the incident electron energy $E$ plotted for several internuclear separations $R_n$ (a.u.).
      The right-hand side of \eqab\ref{eq:boundener} is plotted by the dashed line and the intersections with solid lines determine energies of the \otwom (\dpig) electronic bound state.}
      \label{fig:analqd}
    \end{figure}

    Results of the \rmat calculations discussed here can be used to study effects of the vibronic coupling in the electron collisions with \otwo or in the photodetachment of \otwom.
    That research requires correct characterization of the bound electronic state of \otwom for a range of the relevant nuclear geometries in addition to the scattering phase shifts or analytical quantum defects. \figab\ref{fig:analqd}
    shows the analytical quantum defects for several internuclear separations calculated using Model 3 that provides the most physical results at $R_e$(O$_2$).
    These curves are smooth functions of the incident electron energy, even in the vicinity of the resonance.
    Relatively slow variation of $\mu^0$ with the internuclear separation $R_n$ makes this quantity suitable for modeling the vibrational dynamics of the anionic complex based on the energy-dependent vibrational frame transformation \cite{Gao1990}.
    \figab\ref{fig:analqd} also shows the curve corresponding to the right-hand side of \eqab\eqref{eq:boundener}.
    Its intersection with the smooth quantum defect (if it exists) determines the bound-state energy of \otwom (\dpig).
    
\begin{figure}
      \includegraphics[scale=0.68]{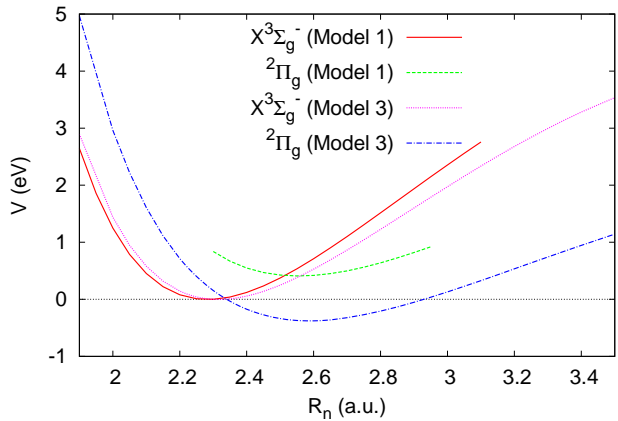}
      \caption{(Color online) Potential energy curves of the target \otwo (\tsigm) ground electronic state and \otwom (\dpig) state.
      In the region where the anion is not stable against autodetachment, the anionic curve displayed represents the real part of the corresponding resonance energy. The calculations using Model 1 and Model 3 are compared. The potential energy curves are plotted for each model with respect to energy of the target ground state at its equilibrium nuclear geometry.
      Model 3 yields a fixed-nucleus electron affinity $E_A=0.375$\thinspace eV.}
      \label{fig:potcurves}
    \end{figure}

    It is well known \cite{Ervin2003, Celotta1972} that \otwom posses one electronic bound state (\dpig) with the minimum of the potential energy curve below the potential energy minimum of \otwo.
    The potential energy curves of the ground state of neutral \otwo and of the anion calculated using Model 1 and Model 3 are plotted in \figab\ref{fig:potcurves}.
    The electronic eigenenergies of the anion are calculated using \eqab\eqref{eq:boundener} for the range of the internuclear separations, where the anion is bound and the resonance energy is taken for smaller internuclear distances, where the anionic state has finite lifetime.
    
    \figab\ref{fig:potcurves} shows that Model 1 does not yield a bound state of the negative ion.
    Although the anionic potential energy curve crosses the the one of the neutral target at $R\approx 2.51$\thinspace a.u., its minimum lies above the minimum of the neutral potential curve.
    This behavior is common for both Models 1 and 2 and can it is explained above.
    The \rmat study by \citet{Higgins1994} also fails to predict the bound state of \otwom.
    The results from the reference \cite{Higgins1994} were later adjusted by shifting the \rmat poles to reproduce the experimental value of the electron affinity and used to study the resonant vibrational excitation of \otwo by the electron impact \cite{Higgins1995}.
    This sort of shift can lead to an inconsistency between the \rmat poles and amplitudes that can nonphysically affect the width of the resonance, and has accordingly not been pursued in the present study.
    
\begin{figure}
      \includegraphics[scale=0.68]{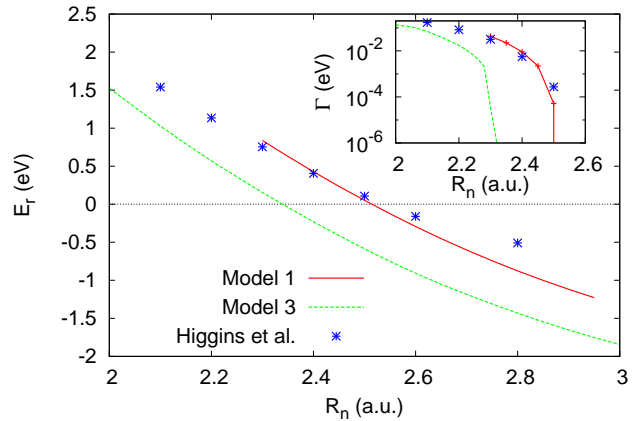}
      \caption{(Color online) The position $E_r$ of the \dpig state of \otwom calculated as a function of the internuclear distance $R$ plotted with respect to the \tsigm electronic threshold.
      The results obtained using Model 1 and Model 3 are compared with previously the published calculations by \citet{Higgins1994}.
      The width of the resonance is compared in the inset.}
      \label{fig:resbound}
    \end{figure}

    The \rmat calculations based on Model 3 that employ the molecular orbitals optimized for the negative ion clearly show the bound state of \otwom and yield the electron affinity $E_A=0.375$\thinspace eV.
    This is in good agreement with the experimental value 0.448\thinspace eV \cite{Celotta1972} supported by later quantum chemical calculation by \citet{Ervin2003} (see also \tabab\ref{tab:spectroscopic}).
    The crossing point of the neutral and anionic potential curve obtained using Model 3 is located at internuclear distance 2.34\thinspace a.u., very close to the equilibrium geometry of the neutral target.
    This position is close to that determined in \citab\cite{Ervin2003}.
    \figab\ref{fig:resbound} shows the resonance and bound-state energy of \otwom relatively to the energy of the neutral target ground state.
    The resonance width as function of $R_n$ is plotted in the inset.
    This visualization allows for a direct comparison of Model 1 and Model 3 with \citab\cite{Higgins1994} for multiple nuclear geometries.
    It confirms that the good agreement between Model 1 and the reference \cite{Higgins1994} is preserved for other nuclear geometries than the equilibrium of the neutral target, while Model 3 yields lower values of the resonance position and width as well as of the energy of the anionic bound state.
    
    The potential energy curves plotted in \figab\ref{fig:potcurves} and the analytical quantum defects plotted in \figab\ref{fig:analqd} are the central physical results of this study, as well as the basis set configurations that seem to produce the best results.
    The resonance and bound-state energies of \otwom presented here show that the \abini \rmat calculations using Model 3 provide \otwo physically correct eigenphases and smooth quantum defects in the range of the collision energies considered here, in spite of the complicated electronic structure, whereas the previously published \abini studies provide only inaccurate results.
    
    Another quantities essential in this context are the harmonic vibrational frequencies of the neutral target and of the bound negative ion.
    As one can see in \tabab\ref{tab:spectroscopic}, the values obtained using Model 1 and Model 3 are in good agreement with the experimental value \cite{Ervin2003}.
  \section{Conclusion}
    \label{ref:conc}
    The \abini study of the electronic structure of the \dpig bound and continuum state of \otwom in the approximation of the fixed nuclei is presented.
     The scattering eigenphases and the analytical quantum defects are given as functions of the scattering energy for the range of the internuclear separations relevant in the resonant vibrational excitation of \otwo and photodetachment of \otwom.
    The scattering energies below 1\thinspace eV, where only one electronic channel is open, are considered.
    For geometries, where the anionic state is not stable against electron autodetachment, the eigenphases were fitted to the Breit-Wigner formula \eqref{eq:bwform} to determine the resonance position and width.
    At larger internuclear distances, where the anion is bound, its energy was determined from the smooth quantum defects using \eqab\eqref{eq:boundener}.
    All the calculations were performed using the UK molecular \rmat program suite \cite{tennyson-rev,Carr2012}.
    
    The results for several different CAS models show that a large basis set of the CI configurations and neutral target eigenstates is necessary in the inner region to achieve converged eigenphases.
    It is found that if the molecular orbitals of the neutral target are employed in the inner region, the convergence of the CAS model with respect to the number of included orbitals is too slow to provide physically correct characterization of the bound and resonant \dpig state of \otwom.
    A different method of selection of the CI configurations other than CAS should be used in that case.
    However, it is not straightforward to find a more appropriate selection scheme for the CI configurations that treats the target and the $(N+1)$-electron scattering complex in a balanced manner \cite{tennyson-rev,Tarana2007}.
    In addition, any such alternative selection scheme would require substantial changes in the existing UK \rmat codes.
    On the other hand, if the molecular orbitals optimized for the negative ion are employed, the CAS approach in the inner region yields the eigenphases and analytical quantum defects that yield the parameters of the resonance and of the bound state in good agreement with the experimental values.
    
    The problem studied here is suitable for a test of the alternative methods of calculation of the \rmat.
    The method tested here is based on the solution of the linear system of equations individually for every scattering energy.
    This method proved more suitable than the single complete diagonalization, when the dimensions of $\underline{H}_\Omega$ exceeds 40000, at least for typical computational hardware that is routinely available at present.
    It represents a nonperturbative alternative to the partitioned \rmat method that requires a single partial diagonalization of the Hamiltonian matrix $\underline{H}_\Omega$ after which the spectrum is completed approximately.
    \begin{acknowledgments}
      This work was supported in part by the Department of Energy, Office of Science. This research used resources of the Janus supercomputer at CU-Boulder.
    \end{acknowledgments}

 \appendix
  \section{Notes on the method used to solve the linear system}
    The method of calculation of the \rmat from the Hamiltonian in the inner region $\underline{H}_\Omega$ introduced in \secab\ref{sec:method} was implemented using the linear solver PARDISO \cite{Schenk2004} that is based on LU factorization.
    This solver employs a sparse matrix storage scheme and it is designed to handle large and sparse matrices that cannot fully fit into the memory using the full storage.
    The UK \rmat program employs the DSYEVD subroutine from the LAPACK library for the complete diagonalization.
    The DSYEVD subroutine requires storage of the full matrix and does not benefit from the fact that $\underline{H}_\Omega$ is usually sparse.
    
    The sparse matrix storage is the main reason, why the method based on solving the system of linear equations is more favorable for large CAS models than the complete diagonalization.
    When the dimension of $\underline{H}_\Omega$ is smaller than $\approx$40000 and when the memory is large enough to store it using the full storage format, the complete diagonalization requires less CPU time than the solution of multiple linear systems for several different scattering energies (see \tabab\ref{tab:laham}).
    Performance of both methods becomes comparable above this dimension and the number of considered scattering energies for which the \rmat needs to be calculated decides which method requires less CPU time.
    This dimension is also close to the memory limit of typical contemporary computers.
    Comparison of both methods beyond this size becomes complicated and considerably larger CAS models can be handled only using methods that do not require full matrix storage.
    
    The CPU time required by PARDISO to solve single system of linear equations also depends on the density of $\underline{H}_\Omega$.
    In general, even very large problem can be solved efficiently using this subroutine, if it is sufficiently sparse.
    
  \section{Comment on the application of the RMPS method}
    \label{sec:aprmps}
    The high number of the target states required to achieve a converged expansion of the total $(N+1)$-electron wave function in the inner region naturally suggests that the approach employing the \rmat with pseudostates (RMPS) method might be rather efficient.
    Previous studies \cite{Gorfinkiel2005,tarana-tennyson,Jones2010,tennyson-rev} show that an accurate treatment of the complicated polarization effects in the inner region requires a smaller number of the pseudostates than of the true eigenstates of the target.
    The RMPS method was applied to the problem studied here and following \citet{Gorfinkiel2005} the singly-excited configurations
    $(1\sigma_g2\sigma_g3\sigma_g1\sigma_u2\sigma_u1\pi_u)^{14}(1\pi_g)^1(\lambda_i)^1$
    were included in the CI expansion of the target states, where $\lambda_i$ is the pseudocontinuum orbital \cite{Gorfinkiel2005}.
    However, those additional configurations do not have any considerable contribution to the eigenstates of the neutral target in any of Models 1--3, where the RMPS method was tested.
    As a consequence, the higher pseudostates do not improve the representation of the polarization effects in the scattering calculations.
    
    As an attempt to improve the representation of the pseudostates, the doubly excited configurations, where the pseudocontinuum orbital is singly occupied, were added.
    These configurations decreased the energies of the target states.
    Corresponding $(N+1)$-electron terms including the pseudocontinuum orbitals were added to the CI expansion of the $(N+1)$-electron wave function as well.
    However, this CI model yields an extremely large Hamiltonian $\underline{H}_\Omega$ with the dimension exceeding $3\times10^5$.
    The evaluation of all the matrix elements would require too much CPU time and the \rmat calculation becomes intractable.
    Therefore, the RMPS treatment of the electron collisions with \otwo at low incident electron energies should be reconsidered and it will be subject of future research.
%
\end{document}